\documentclass[12pt]{article}
\usepackage{times,amssymb,euscript,epsfig,latexsym,amsmath}
\usepackage{graphicx}
\newtheorem{thm}{Theorem}[section]

\newtheorem{dfn}[thm]{Definition}

\date{}
\title{Vortex dynamics in $\mathbb{R}^4$}
\author{Banavara N. Shashikanth\footnote{Mechanical and Aerospace Engineering
Department, MSC 3450, PO Box 30001,
New Mexico State University, Las Cruces, NM 88003, USA.
E-mail:shashi@nmsu.edu, ph:575-646-4348 }}
\begin{document}
\maketitle
\begin{abstract}
The vortex dynamics of Euler's equations for a constant density fluid flow in $\mathbb{R}^4$ is studied. Most of the paper focuses on singular Dirac delta distributions of the vorticity two-form $\omega$ in $\mathbb{R}^4$. These distributions are supported on two-dimensional surfaces termed {\it membranes} and are the analogs of vortex filaments in $\mathbb{R}^3$ and point vortices in $\mathbb{R}^2$. The self-induced velocity field of a membrane is shown to be unbounded and is regularized using a local induction approximation (LIA). The regularized self-induced velocity field is then shown to be proportional to the mean curvature vector field of the membrane but rotated by 90$^{\circ}$ in the plane of normals. Next, the Hamiltonian membrane model is presented. The symplectic structure for this model is derived from a general formula for vorticity distributions due to Marsden and Weinstein  \cite{MaWe83}. Finally, the dynamics of the four-form $\omega \wedge \omega$ is examined. It is shown that Ertel's vorticity theorem in $\mathbb{R}^3$, for the constant density case, can be viewed as a special case of the dynamics of this four-form.
\end{abstract}
\tableofcontents

\section{Introduction.}
The  Navier-Stokes and Euler equations for a classical fluid in a physical domain of spatial
dimensions greater than three have typically attracted mathematical investigations aimed at understanding fundamental issues
 related to the equations and their solutions. On an $n$-dimensional Riemannian manifold, for example, these
 equations  have been investigated for their deep geometric, topological and functional analytic
 properties---good introductions to such research with many important references can be found in
 the books by Arnold and Khesin \cite{ArKh98} and Temam \cite{Te95}.
There is also a slender body of theoretical and computational work---the latter no doubt facilitated by major advances in computational power in the last decade---on  the Navier-Stokes equations in higher dimensions with the focus on turbulence.  A few examples of such work are \cite{SuNaTaGo05, Ni11, FoFrRo78, GoWaShNaSu07, FoFr78}.

         It is fair to say that, in general, classical fluid flows governed by these equations in higher dimensions--even if one restricts to $\mathbb{R}^4$--have been far less investigated than flows in $\mathbb{R}^3$ and $\mathbb{R}^2$. In particular, little attention seems to have been paid to the development of low-(phase-space)-dimensional mathematical models of flows in $\mathbb{R}^4$. Motivated by this, this paper focuses on a set of mathematical models of vorticity, in the framework of Euler's equations, which have found much favor amongst  mathematicians, physicists and engineers for studying flows in $\mathbb{R}^2$ and $\mathbb{R}^3$. These are models of the dynamics and evolution of coherent vortices in which the vorticity field is a singular Dirac delta field. Examples of such models are point vortices, vortex filaments/rings and vortex sheets. A good introduction to these models can be found in Saffman \cite{Sa92}.

         The importance and significance of vorticity for flows in three or less dimensions is so well-established that it needs no elaboration. It is therefore to be expected that it plays an equally important role for flows in higher dimensions. Vortices, as is well-known, also occur in other areas of physics most notably in superfluids. For this paper, however, the focus will be entirely on classical fluids.

           The outline of the paper is as follows. In section 2, some basic facts of the vorticity two-form in $\mathbb{R}^4$ are presented. In section 3, the notion of a vortex membrane in $\mathbb{R}^4$ is introduced. A membrane is a singular Dirac delta distribution of vorticity supported on a two-dimensional surface. It is shown that the self-induced velocity field of a membrane is unbounded and a local induction approximation (LIA) is used to regularize the velocity. In section 4, the Hamiltonian membrane model is introduced. It is shown that the evolution equation for a membrane---again, after a LIA regularization---has Hamiltonian structure relative to a symplectic form derived from a general formula for vorticity distributions due to Marsden and Weinstein \cite{MaWe83}.  In section 5, the relation of the (regularized) velocity field to the mean curvature vector field of the membrane is made clear. In section 6, the dynamics of the four-form $\omega \wedge \omega$---which is identically zero in $\mathbb{R}^3$ and lower dimensions---is examined and it is shown that Ertel's vorticity theorem in $\mathbb{R}^3$, for the constant density case, can be viewed as a special case of the evolution equation of this four-form.
\section{Basics of vorticity in $\mathbb{R}^4$.}
 In $\mathbb{R}^4$, the vorticity field is a genuinely two-form field. It can no longer be associated with a vector field, as in $\mathbb{R}^3$, or with  a scalar field, as in $\mathbb{R}^2$, but can be represented, if necessary, as a skew-symmetric matrix field. It follows that a natural setting for investigating vorticity in higher dimensions is the exterior algebraic/differential geometric framework, along with the associated Hamiltonian formalism, pioneered by Arnold \cite{Ar66, Ar69} and developed, among others, by Marsden and Weinstein \cite{MaWe83}.

 Equipping $\mathbb{R}^4$ with the Euclidean metric, the 1-form associated with the velocity field (through the metric) $v(x_i,t)=(v_i(x_i,t)), \: i=1,\cdots,4$, is
\begin{align}
	v^{\flat}&= \sum_{i=1}^4 v_i dx_i
\end{align}
satisfying the divergence-free condition
\begin{align}
	\operatorname{Div}(v)&= -\eth v^{\flat}=\ast {\bf d} \ast v^{\flat}=0,
\end{align}
where $\eth: \Omega^{k+1}  (\mathbb{R}^4) \rightarrow \Omega^{k}  (\mathbb{R}^4)$ is the co-differential operator and $\ast:\Omega^{k}  (\mathbb{R}^4) \rightarrow \Omega^{4-k}  (\mathbb{R}^4)$ is the Hodge star operator.\footnote{For the minus sign in front of $\eth$, see \cite{AbMaRa88}, page 458} Here, $\Omega^{k}$ represents the space of differential $k$-forms on $\mathbb{R}^4$.

The vorticity two-form is
\begin{align}
	\omega&= {\bf d}v^{\flat} \\
	&= \left(-\frac{\partial v_1}{\partial x_2} + \frac{\partial v_2}{\partial x_1} \right) dx_1 \wedge dx_2 + \left( -\frac{\partial v_1}{\partial x_3} + \frac{\partial v_3}{\partial x_1}\right) dx_1 \wedge dx_3 + \left( -\frac{\partial v_1}{\partial x_4} + \frac{\partial v_4}{\partial x_1} \right) dx_1 \wedge dx_4  \nonumber \\
& + \left(-\frac{\partial v_2}{\partial x_3} + \frac{\partial v_3}{\partial x_2} \right) dx_2 \wedge dx_3 + \left( -\frac{\partial v_2}{\partial x_4} + \frac{\partial v_4}{\partial x_2}\right) dx_2 \wedge dx_4  + \left( -\frac{\partial v_3}{\partial x_4} + \frac{\partial v_4}{\partial x_3}\right) dx_3 \wedge dx_4  \nonumber \\
	&=: \omega_{12} dx_1 \wedge dx_2 + \omega_{13} dx_1 \wedge dx_3 + \omega_{14} dx_1 \wedge dx_4  \nonumber \\
	& \hspace{0.5in} + \omega_{23} dx_2 \wedge dx_3 + \omega_{24} dx_2 \wedge dx_4 +  +\omega_{34} dx_3 \wedge dx_4  \label{eq:vort2form}
\end{align}

\paragraph{Closedness of $\omega$.} Since $\omega$ is exact, it is closed i.e ${\bf d} \omega=0$. However, there seems to be no obvious way of identifying ${\bf d} \omega=0$ with a divergence-free condition as in $\mathbb{R}^3$ or $\mathbb{R}^2$. In the above coordinates, the condition gives rise to the equation
\begin{align}
&	\left( \frac{\partial \omega_{12} }{\partial x_3} - \frac{\partial \omega_{13} }{\partial x_2} + \frac{\partial \omega_{23} }{\partial x_1} \right)dx_1 \wedge dx_2 \wedge dx_3 + \left( \frac{\partial \omega_{12} }{\partial x_4} - \frac{\partial \omega_{14} }{\partial x_2} + \frac{\partial \omega_{24} }{\partial x_1} \right)dx_1 \wedge dx_2 \wedge dx_4 \\
&+ \left( \frac{\partial \omega_{13} }{\partial x_4} - \frac{\partial \omega_{14} }{\partial x_3} + \frac{\partial \omega_{34} }{\partial x_1} \right)dx_1 \wedge dx_3 \wedge dx_4 + \left( \frac{\partial \omega_{23} }{\partial x_4} - \frac{\partial \omega_{24} }{\partial x_3} + \frac{\partial \omega_{34} }{\partial x_2} \right)dx_2 \wedge dx_3 \wedge dx_4=0
\end{align}
at all points in $\mathbb{R}^4$. This gives rise to four equations which can be written in compact form as
\begin{align}
  \frac{\partial \omega_{ij}}{\partial x_k}&=0, \quad (i,j,k)=\left\{(1,2,3),(1,2,4),(1,3,4),(2,3,4) \right \} \label{eq:divfree}
\end{align}
with the convention that $\omega_{ki}=-\omega_{ik}$ if $k > i$ and so on.

The Lie-Poisson equation for vorticity evolution \cite{MaWe83} is
\begin{align}
	\frac{\partial \omega}{\partial t}+ \mathcal{L}_v \omega&=0 \label{eq:lpv}
\end{align}
Using exterior algebraic properties of Lie derivatives \cite{AbMaRa88}, it can be shown that
the equations of each component of $\omega$---in coordinates---are
\begin{align}
	\frac{\partial \omega_{ij}}{\partial t}+ \sum_{k=1,k \neq j }^4 \frac{\partial (\omega_{kj} v_k)}{\partial x_i} + \sum_{k=1,k \neq i}^4 \frac{\partial (\omega_{ik} v_k)}{\partial x_j}&=0, \quad j > i \label{eq:form1}
\end{align}
In the above form of writing the equations, the convention adopted is $\omega_{kj}=-\omega_{jk}$ if $k >j$ and $\omega_{ik}=-\omega_{ki}$ if $i >k$.

     \paragraph{Vortex surfaces.} The essentially two-form nature of $\omega$ in $\mathbb{R}^4$ means that there are no vortex lines, i.e. curves tangent to vorticity vector fields \cite{Sa92} since the latter do not exist. From an exterior algebraic viewpoint, a vorticity vector field in $\mathbb{R}^3$ is the vector field $(\ast \omega)^{\sharp}$, where $\ast$ is the Hodge star operator and $\sharp$ is the operator that relates 1-forms to vector fields via the metric. In $\mathbb{R}^4$, however, $\ast \omega$ is a 2-form and there is no canonical way of identifying $(\ast \omega)^{\sharp}$ with a vector field.

     To elaborate, consider, more generally, (local) submanifolds $\mathcal{S} \subset \mathbb{R}^n$ to which  the two-form $\omega$ is {\it perpendicular} or, equivalently,
  \begin{align}
	i_{\mathcal{S}}^* \omega &=0, \label{eq:perpform}
\end{align}
 where $i_\mathcal{S}: \mathcal{S} \rightarrow \mathbb{R}^n$ is the inclusion map and superscript $*$ denotes pullback of forms. Since $\ast \ast \omega=(-1)^{2(n-2)} \omega$, the above equation is equivalent to saying that $\ast \omega$ is {\it tangent} to $\mathcal{S}$ (see \cite{AbMaRa88}, p.540,  for an introduction to perpendicular and tangent forms). For $\ast \omega$ to be a non-trivial form on $\mathcal{S}$, it follows that the dimension of $\mathcal{S}$ must be equal to or greater than the order of the form i.e. $\operatorname{dim}(\mathcal{S}) \ge n-2$. The lowest possible dimension of $\mathcal{S}$ is therefore one\footnote{And zero in $\mathbb{R}^2$ consistent with the fact that vorticity in the plane is equivalent to a scalar function. The above geometric argument also shows why `vortons' in $\mathbb{R}^3$ are not genuine representations of vorticity; for more on vortons see, for example, Novikov \cite{No83} or Winckelmans and Leonard \cite{WiLe93}} in $\mathbb{R}^3$  and
 two in $\mathbb{R}^4$.
 Hence, in $\mathbb{R}^4$ there are no vortex lines---instead, one thinks of {\it vortex surfaces} which satisfy~(\ref{eq:perpform}) at each point.

\section{Vortex membranes and 3-branes.}

\paragraph{Singular Dirac delta vorticity fields: vortex membranes and 3-branes} Recall that in $\mathbb{R}^2$, when the voriticy is a singular Dirac delta field supported on a $\mathcal{S}$ with $(\operatorname{Dim(\mathcal{S})}=0,\operatorname{Dim(\mathcal{S})}=1)$ one obtains the (point vortex, vortex sheet) models. Similarly, in $\mathbb{R}^3$, when the voriticy is a singular Dirac delta field supported on a $\mathcal{S}$ with $(\operatorname{Dim(\mathcal{S})}=1,\operatorname{Dim(\mathcal{S})}=2)$ one obtains the (vortex filament, vortex sheet) models.\footnote{The $\mathbb{R}^2$ models are nothing but special cases of the $\mathbb{R}^3$ models and it is sometimes useful to adopt this viewpoint.}

Singular Dirac delta  vorticity fields in $\mathbb{R}^4$ that are the analogs of point vortices
and vortex filaments in $\mathbb{R}^2$ and $\mathbb{R}^3$, respectively, are two-dimensional surfaces
$\Sigma$---henceforth referred to as {\it membranes}---such that
 \[i_{\Sigma}^* \omega=0, \]
 at all points of $\Sigma$, where $i:\Sigma \rightarrow \mathbb{R}^4$ is the inclusion. The precise sense in which the vorticity field is a Dirac delta field supported on $\Sigma$ is made clear further down.


    Note that a vortex sheet in $\mathbb{R}^3$, when embedded in $\mathbb{R}^4$, is not a vortex membrane because the vorticity two-form acts on different planes for a sheet and a membrane. For a sheet embedded in $\mathbb{R}^4$, the vorticity two-form  at any point acts in a plane that  intersects the sheet non-transversally in $\mathbb{R}^4$. But for a membrane the vorticity two-form  at any point acts in a plane that  intersects the membrane transversally in $\mathbb{R}^4$.

      Just as vortex filaments in $\mathbb{R}^3$ can be `stacked together' in such a manner  that the vorticity two-form continues to be perpendicular to the resultant surface, to form a vortex sheet,  similarly, membranes in $\mathbb{R}^4$ can be stacked together so that the  vorticity two-form continues to be perpendicular to the resultant volume to give analogs of vortex sheets which are termed {\it 3-branes}.

For the rest of this paper, however, the focus will be only on membranes.

  \paragraph{Vorticity two-form for a membrane.}  To write the vorticity two-form for a membrane in an appropriate basis, recall an elementary fact about 2D surfaces in $\mathbb{R}^4$. Unlike in $\mathbb{R}^3$ where the normal vector is defined uniquely at any point of a 2D surface, there is no unique vector at each point of the surface but rather {\it a plane of normal vectors.} At any point $p \in \Sigma$, the local decomposition thus exists
\begin{align*}
    T_p \mathbb{R}^4&= T_p \Sigma \oplus N_p,
\end{align*}
where $N_p$ is the span of the normal vectors  to $\Sigma$ at $p$. And so the most natural way to write the vorticity two-form, which is perpendicular to $\Sigma$, is as a volume form on $N_p$---similar to how Marsden and Weinstein \cite{MaWe83} wrote it for vortex filaments in $\mathbb{R}^3$. As they noted, the form ${\bf i}_{t} \mu$ is a volume form on the plane of normals at each point of a curve in $\mathbb{R}^3$, where $\mu$ is the volume form on $\mathbb{R}^3$, $t$ is the unit tangent vector to the curve  and ${\bf i}$ is the interior product \cite{AbMaRa88}. To see this, simply choose an orthonormal basis $(t,n_1,n_2)$ for $\mathbb{R}^3$ at each point of the curve, where $(n_1,n_2)$ is a basis  for the plane of normals, to obtain ${\bf i}_{t} \mu=dn_1 \wedge dn_2$

 Similarly, for membranes choose an orthonormal basis $(n_1,n_2, t_1,t_2)$ for $\mathbb{R}^4$ at each $p \in \Sigma$, where $n_1,n_2 \in N_p$ and $(t_1,t_2) \in T_p \Sigma$.  The volume form on $\mathbb{R}^4$ in this basis\footnote{The volume form can also be written in the canonical basis  $\mu= dx_1 \wedge dx_2 \wedge dx_3 \wedge dx_4$ and noting that $dx_1 \wedge dx_2 \wedge dx_3 \wedge dx_4=dn_1 \wedge dn_2 \wedge dt_1 \wedge dt_2$ since the two bases are related by an orthogonal transformation.} is
 \begin{align}
   \mu&= dn_1 \wedge dn_2 \wedge dt_1 \wedge dt_2 \label{eq:stdvolumeform}
   \end{align}
   and one obtains
    \begin{align}
    {\bf i}_{t_1} {\bf i}_{t_2} \mu=dn_1 \wedge dn_2 \label{eq:volumeform}
    \end{align}
     as a volume form on $N_p$.

      It follows that the vorticity two-form for a single vortex membrane of strength $\Gamma$ in $\mathbb{R}^4$ can be written as
  \begin{align}
   \omega(m)&=\Gamma \delta(m-p){\bf i}_{t_1} ({\bf i}_{t_2} \mu), \quad m \in \mathbb{R}^4, \: p \in \Sigma,  \nonumber \\
   &=\Gamma \delta (m-p) dn_1 \wedge dn_2, \label{eq:vortmembranes}
   \end{align}
   where $\delta(m-p)$ is the Dirac delta function supported on $\Sigma$.

     Note  that membranes in $\mathbb{R}^4$, like filaments in $\mathbb{R}^3$, are boundaryless due to the closedness of the vorticty two-form. Co-dimension two vortex surfaces, such as membranes,  have also been examined in the context of the Ginzburg-Landau equations by, for example, Lin \cite{Li98} and Jian and Liu \cite{JiLi06}.

   \paragraph{Assumption.} For the rest of the paper, the membranes considered will be assumed to be compact and arcwise-connected subsets of $\mathbb{R}^4$. The former assumption is made simply to avoid technicalities of Stokes' theorem for non-compact manifolds and it is conjectured that most of the results that follow also hold for non-compact membranes. The latter assumption is needed for the proof in Appendix~\ref{strengthindep}. From a physical point of view, neither of these assumptions are considered restrictive.

  \paragraph{Strength of a membrane.} Before proceeding further, the notion of the strength of a membrane, $\Gamma$, needs to be defined and clarified.
   \begin{dfn}\label{strength}The strength of a vortex membrane $\Sigma$  is defined as
   \begin{align*}
    \Gamma&:= \int_{A} i_{A}^* \omega,
   \end{align*}
   where $A$ is a smooth two-dimensional disc-like  surface  which intersects $\Sigma$ transversally at a single point, such that $\partial A$ (a closed curve) cannot be shrunk to a point without intersecting $\Sigma$, and $i_{A}:A \rightarrow \mathbb{R}^4$ is the inclusion map. Equivalently, using Stokes theorem and since $\omega={\bf d}v^{\flat}$,
   \begin{align}
    \Gamma&:= \oint_{\partial A} \tilde{i}_{\partial A}^* i_{A}^* v^{\flat} \equiv \oint_{\partial A} i_{\partial A}^*v^{\flat}, \label{eq:circulation}
   \end{align}
   where $\tilde{i}_{\partial A}: \partial A \rightarrow A$ is the inclusion and $i_{\partial A}:=i_A \circ \tilde{i}_{\partial A}$.
   \end{dfn}
   \paragraph{Notes:}\begin{itemize}
   \item[1.]  Since the intersection is transversal,  $\Sigma$, which is of co-dimension two, intersects $A$, also of co-dimension two, only at isolated points.
   \item[2.] As in $\mathbb{R}^3$, the definition using the notion of circulation,~(\ref{eq:circulation}), can be taken as the basic definition of $\Gamma$ and the other definition, using vorticity, can be viewed as a consequence of Stokes theorem.
    \item[3.] Since $\omega$ is a Dirac delta field, the integral in the above definition should be interpreted---like in $\mathbb{R}^3$----as the limiting value of the integral defined for a tubular neighborhood of the membrane. This is made clear below.
        \item[4.] For a given $\partial A$ there can be many $A$s but, as in $\mathbb{R}^3$, Stokes theorem applied to $A$ or, equivalently,~(\ref{eq:circulation}) shows that the choice of $A$ is irrelevant.
        \end{itemize}
        The next step is to show that the above definition is independent of  the choice of  $\partial A$. The proof of this is presented in Appendix~\ref{strengthindep}.

   \subsection{Self-induced velocity field of a membrane.} In this subsection, the self-induced velocity field of a membrane is examined. It is shown that---as for filaments in $\mathbb{R}^3$---the field diverges and requires a regularization. A local induction approximation (LIA) is then used to show that the leading order term in the regularized field is proportional to the {\it mean curvature vector of $\Sigma$ rotated by 90 degrees in the plane of normals}   at each point.

   To show this, first express the velocity field of a given smooth vorticity field. The relationship $\nabla^2 v=-\nabla \times \omega$ generalizes in $\mathbb{R}^4$ as
   \begin{align*}
    \eth \omega&=\eth {\bf d}v^{\flat},
   \end{align*}
   where $\eth$ is the co-differential operator. The above is obtained simply by applying the co-differential operator to both sides of the definition of the vorticity two-form. The operator $\eth {\bf d}$ on functions is nothing but the Laplacian in $\mathbb{R}^4$. Writing $v^{\flat}= \sum_{i=1}^4 v_i \wedge dx_i$, where $v_i: \mathbb{R}^4 \rightarrow \mathbb{R}$, obtain
   \begin{align*}
    \eth \omega&=\eth {\bf d} \sum_{i=1}^4 (v_i \wedge dx_i), \\
    &= (\nabla^2 v)^{\flat}.
   \end{align*}
   The last line can be obtained by either expanding the terms on the right side of the previous line in coordinates or using properties of the exterior algebra objects (wedge product, ${\bf d}$, $\eth$ and $\ast$). Note that use is made of the divergence-free property of $v$. The equation is now equivalent to the Poisson equation
   \begin{align*}
    \nabla^2 v_i&= \ast (dx_i \wedge \ast  \eth \omega  ), \quad i=1, \cdots,4
   \end{align*}
   Denoting by $G(m,m_0)$ the Green's function of the Laplacian in $\mathbb{R}^4$, where $m,m_0 \in \mathbb{R}^4$, the solution to the above is
   \begin{align}
    v_i(m) &= \int_{\mathbb{R}^4} G(m,m_0) \wedge \ast (dx_i \wedge \ast  \eth \omega  ) \; \mu_{m_0}, \quad i=1, \cdots,4 \nonumber  \\
    &=- \int_{\mathbb{R}^4} \left< \left< e_i , (\ast ({\bf d}G(m,m_0)  \wedge \ast  \omega))^{\sharp} \right> \right> \mu_{m_0}, \quad i=1, \cdots,4  \label{eq:velint}
   \end{align}
   using properties of the exterior algebra objects and integration by parts. To apply the latter, it is of course assumed that $\omega$ decays sufficiently rapidly at infinity for the boundary term in integration by parts to vanish. In the above, $\mu_{m_0}$ is the canonical volume form on $\mathbb{R}^4$ where the subscript $m_0$ means that the variable of integration is $m_0$, and $\left<\left<;,;\right>\right>$ is the standard metric on $\mathbb{R}^4$. Note that  the variable in the ${\bf d}$ operation is again $m_0$.

      To obtain the self-induced velocity field of $\Sigma$, substitute~(\ref{eq:vortmembranes}) in~(\ref{eq:velint}) and switch the $\mathbb{R}^4$ basis from the canonical to $(n_1,n_2,t_1,t_2)$ to get
        \begin{align}
    v_{SI}(m)&= - \Gamma \int_{\mathbb{R}^4}  \delta(m-p) (\ast ({\bf d}G(m,p)  \wedge (dt_1 \wedge dt_2)))^{\sharp} \mu_{p}, \quad m \in \mathbb{R}^4, p \in \Sigma \nonumber \\
    &= - \Gamma \int_{\mathbb{R}^4}  \delta(m-p) (\frac{\partial G}{\partial n_1} dn_2-\frac{\partial G}{\partial n_2} dn_1)^{\sharp} \mu_{p},  \nonumber \\
    &=  \Gamma \int_{\mathbb{R}^4}  \delta(m-p)( (\nabla  G \cdot n_2)  n_1-(\nabla  G \cdot n_1)  n_2) \mu_{p}, \nonumber  \\
    &=  \Gamma \int_{\Sigma}   ((\nabla  G \cdot n_2)  n_1-(\nabla  G \cdot n_1)  n_2) \nu_p, \quad m, p \in \Sigma \label{eq:selfvelint}
   \end{align}
   In the above,  $\mu_p$ is the same as~(\ref{eq:stdvolumeform}) and $\nu_p=dt_1 \wedge dt_2$.

 \paragraph{Remark.} \label{remark1} There is no canonical way of fixing $n_1$ and $n_2$ or $t_1$ and $t_2$.  The velocity expression should however be invariant to  the choice of the normals and the tangents.  But this is a straightforward check. The integrand in~(\ref{eq:selfvelint}) and $\nu_p$ are invariant under  the (orthogonal) transformation in the plane of normals $(n_1,n_2,t_1,t_2) \rightarrow (\tilde{n}_1,\tilde{n}_2,t_1,t_2)$ and the (orthogonal) transformation in the tangent plane $(n_1,n_2,t_1,t_2) \rightarrow (n_1,n_2,\tilde{t}_1,\tilde{t}_2)$. \\

 Recall that in $\mathbb{R}^4$,
     \begin{align}
     G(m,p)&= - \frac{1}{\pi^2 (\mid r-\tilde{r} \mid^2)}, \label{eq:greens}
     \end{align}
     where $r$ and $\tilde{r}$ are the position vectors of points $m$ and $p$, respectively. And so,
     \begin{align}
        \nabla G&= -\frac{2 (r-\tilde{r})}{ \pi^2 (\mid r-\tilde{r} \mid^4)} \label{eq:gradG}
     \end{align}
     (gradient with respect to $\tilde{r}$). Substituting~(\ref{eq:gradG}) in~(\ref{eq:selfvelint}), it is obvious that the integrand develops a singularity as $p \rightarrow m$. The integral~(\ref{eq:selfvelint}) therefore needs to be evaluated as an improper integral. More specifically, the behavior of the leading order term of the integral as $p \rightarrow m$ is needed. Since there is no divergence in the integral due to non-local effects (as can be easily seen by noting that $(\nabla G \cdot n_i)n_j \nu_p=O(1/\tilde{r})$ as $\tilde{r} \rightarrow \infty$), it suffices to evaluate the integral in a neighborhood of $m$.

        Consider two  closed neighborhoods $U_\delta, U_\epsilon \subset \Sigma$ of $m$, such that $U_\epsilon \subset U_\delta$. The parameters $\delta$ and $\epsilon$ measure the size of these sets (discussed more below). Let $U:=U_\delta \backslash U_\epsilon$ denote the resulting annular domain and
 \begin{align*}
    I&:=\Gamma \int_{U}   ((\nabla  G \cdot n_2)  n_1-(\nabla  G \cdot n_1)  n_2) \nu_p
 \end{align*}
 Proceed now to examine the limit
 \begin{align}
   {\rm Leading \; order \; term \; of \;} v_{SI}(m):= \lim_{U_\epsilon \rightarrow m} I, \label{eq:limit}
 \end{align}
 with $U_\delta$ fixed.

 \paragraph{Coordinates.} All the integrals presented above are integrals of differential forms on smooth manifolds and by their definition such integrals are evaluated using coordinate charts or, in other words, mapping the domain of integration smoothly to/from $\mathbb{R}^n$. The value of the integral is of course independent of the choice of the coordinates but a judicious choice simplifies the evaluation. In particular, a set of coordinates with the following two properties are desirable: (i) the level curves of the coordinates intersect orthogonally and (ii) the coordinates denote arc-lengths.

    Consider the following system of curvilinear polar coordinates consisting of smooth  curves parametrized by arclength $s_1$ measured (in the $\mathbb{R}^4$ metric)
    from $m$ and all passing through $m$.
 Call these `diameter' curves and these do not meet except at $m$. Each of them connects two points  on the boundary of $U_\delta$.
  Similarly, consider a set of closed curves enclosing $m$  parametrized by arclength $s_2$
 such that each closed curve intersects every diameter curve orthogonally at exactly the same length of the diameter curve from $m$.  To each point in $U_\delta$ is thus associated a unique pair of coordinates $(s_1,s_2) \in \mathbb{R}^2$ and this defines a coordinate map $\phi: (s_1,s_2) \rightarrow p$.  Define also a coordinate $\theta$---the angle made between the tangent vector to a `radial' curve at $m$ and any fixed line in $T_{m} \Sigma$---such that $s_2=f(s_1,\theta)$, where $f$ is a smooth (well-defined) function with $f(s_1,0)=0$.  $\theta$ is associated with the entire `diameter' curve.

And so $I$ is evaluated as
 \begin{align*}
    I&=\Gamma \int_{U}   \phi^*(((\nabla  G \cdot n_2)  n_1-(\nabla  G \cdot n_1)  n_2)) \phi^*(\nu_p)
 \end{align*}
 The pullback of the integrand is simply the integrand composed with $\phi$. To evaluate $\phi^*(\nu_p)$, first note that since the choice of $(t_1, t_2)$ is arbitrary, fix these vectors by taking them to be tangent to the level curves of $s_1$ and $\theta$ at each point of $U_\delta$.   Next, write down the coordinate map explicitly using the $\mathbb{R}^4$ coordinates of $p$
 \begin{align}
 \phi:(s_1,s_2)\mapsto (x(s_1,s_2),y(s_1,s_2),z(s_1,s_2),o(s_1,s_2))   \label{eq:chart}
 \end{align}
 and a straightforward computation, using properties (i) and (ii) of the coordinates, gives $\phi^* (\nu_p)=ds_1 \wedge ds_2 \equiv ds_1 ds_2$. And so
 \begin{align}
     I&=\Gamma \int_{f(\epsilon,2\pi)}^{f(\delta,2\pi)} \int_{\epsilon}^\delta   ((\nabla  G \cdot n_2)  n_1-(\nabla  G \cdot n_1)  n_2)(\phi(s_1,s_2)) ds_1 ds_2, \nonumber \\
     & \equiv \Gamma  \int_{0}^{2\pi} \int_{\epsilon}^\delta  ((\nabla  G \cdot n_2)  n_1-(\nabla  G \cdot n_1)  n_2)(\phi(s_1,f(s_1,\theta))) g(s_1,\theta) ds_1 d\theta, \label{eq:inted}
\end{align}
 where $g(s_1,\theta)=\partial f(s_1,\theta) / \partial \theta$. Since the level curves of $s_1$ shrink to zero as $s_1 \rightarrow 0$, one observes that  $g(s_1,\theta) \rightarrow 0$ as $s_1 \rightarrow 0$ (for all $\theta$).

%
%

  At each point in $U_\delta$, there exists an orthogonal frame $(n_1(s_1,\theta),n_2(s_1,\theta),t_1(s_1,\theta),t_2(s_1,\theta))$ and, in particular, at the point $m$, there exists the one-parameter set of frames $(n_1(0),n_2(0),t_1(0,\theta),t_2(0,\theta))$ tangent to the `diameter' curves.  Wlog, let $t_1(0,0)=t_1$ and $t_1(0,\pi/2)=t_2$. Taking $r=0$, expand $\tilde{r}$ for small $s_1$ and fixed $\theta$ along each `diameter' curve
     \begin{align*}
        \tilde{r}(s_1,\theta)&= s_1 \frac{\partial \tilde{r}}{\partial s_1}(0,\theta) + \frac{s_1^2}{2} \frac{\partial^2 \tilde{r}}{\partial s_1^2}(0,\theta) + \frac{s_1^3}{6} \frac{\partial^3 \tilde{r}}{ds_1^3} (0,\theta)+ \frac{s_1^4}{24} \frac{\partial^4 \tilde{r}}{\partial s_1^4}(0,\theta)+O(s_1^5), \\
        &= s_1 t_1(0,\theta) + \frac{s_1^2}{2} \frac{\partial t_1}{\partial s_1} (0,\theta) + \frac{s_1^3}{6} \frac{\partial^2 t_1}{\partial s_1^2}(0,\theta)+ \frac{s_1^4}{24} \frac{\partial ^3 t_1}{\partial s_1^3}(0,\theta)+O(s_1^5),
     \end{align*}
      Similarly, for $i=1,2$,
     \begin{align*}
       & n_i(s_1,\theta) \\
       &=n_i(0)+ s_1 \frac{\partial n_i }{\partial s_1}(0,\theta) + \frac{s_1^2}{2} \frac{\partial^2 n_i}{\partial s_1^2}(0,\theta) + \frac{s_1^3}{6} \frac{\partial^3 n_i}{ds_1^3} (0,\theta)+ \frac{s_1^4}{24} \frac{\partial^4 n_i}{\partial s_1^4}(0,\theta)+O(s_1^5),
     \end{align*}
     and
     \begin{align*}
       & g(s_1,\theta) \\
       &= s_1 \frac{\partial g }{\partial s_1}(0,\theta) + \frac{s_1^2}{2} \frac{\partial^2 g}{\partial s_1^2}(0,\theta) + \frac{s_1^3}{6} \frac{\partial^3 g}{ds_1^3} (0,\theta)+ \frac{s_1^4}{24} \frac{\partial^4 g}{\partial s_1^4}(0,\theta)+O(s_1^5),
     \end{align*}
     For small $s_1$, $f(s_1,\theta) \sim s_1 \theta$ since the  mesh of coordinate curves must approach a mesh of polar coordinates in the tangent plane at $m$. And so set $\partial g (0,\theta)/ \partial s_1=1$ in the above series expansion.

     Using~(\ref{eq:gradG}), the integrand of $I$ (modulo the constant factor $2\Gamma / \pi^2$) is
     \[\frac{(\tilde{r} \cdot n_2)n_1- (\tilde{r} \cdot n_1)n_2}{\mid \tilde{r} \mid^4}(s_1,\theta) g(s_1,\theta), \]
     Substituting the series expansions for small $s_1$, (suppressing the $(s_1,\theta)$ notation temporarily)
     \begin{align*}
     \frac{1}{\mid \tilde{r} \mid^4}&= \frac{1}{\left(s_1^2 + s_1^3 t_1 \cdot \frac{\partial t_1}{\partial s_1} + O(s_1^4) \right)^2}, \\
     &= \frac{1}{s_1^4} \left(1 -2  s_1 t_1 \cdot \frac{\partial t_1}{\partial s_1}  + O(s_1^2) \right)
     \end{align*}
     Next,
     \begin{align*}
        &(\tilde{r} \cdot n_2)n_1- (\tilde{r} \cdot n_1)n_2 \\
        &= s_1^2 \left[\left(t_1 \cdot \frac{\partial n_2}{\partial s_1} + \frac{1}{2} n_2(0)  \cdot \frac{\partial t_1}{\partial s_1}\right)n_1(0) -\left( t_1 \cdot \frac{\partial n_1}{\partial s_1} + \frac{1}{2} n_1(0)  \cdot \frac{\partial t_1}{\partial s_1}\right)n_2(0)  \right] + O(s_1^3)
     \end{align*}
     so that (re-introducing the $(s_1,\theta)$ notation),
     \begin{align*}
        & \frac{(\tilde{r} \cdot n_2)n_1- (\tilde{r} \cdot n_1)n_2}{\mid \tilde{r} \mid^4}(s_1,\theta)g(s_1,\theta) \\
        &= \frac{1}{s_1} \left[\left(t_1(0,\theta) \cdot \frac{\partial n_2(0)}{\partial s_1} + \frac{1}{2} n_2(0)  \cdot \frac{\partial t_1(0,\theta)}{\partial s_1}\right)n_1(0)  \right. \\
        & \hspace{1in}-\left. \left( t_1(0,\theta) \cdot \frac{\partial n_1(0)}{\partial s_1} + \frac{1}{2} n_1(0)  \cdot \frac{\partial t_1(0,\theta)}{\partial s_1}\right)n_2(0)  \right] + O(1), \\ \\
        &=\frac{1}{s_1} \left[\left( \frac{1}{2} t_1(0,\theta) \cdot \frac{\partial n_2(0)}{\partial s_1} \right)n_1(0) - \left( \frac{1}{2} t_1(0,\theta) \cdot \frac{\partial n_1(0)}{\partial s_1}\right)n_2(0)  \right] + O(1),
     \end{align*}
     using the fact that $t_1(s,\theta) \cdot n_i(s_1,\theta)=0$ $\forall (s_1,\theta)$.
     Substituting in~(\ref{eq:inted}) and evaluating the limit~(\ref{eq:limit}) one obtains
     \begin{align}
       &  {\rm Leading \; order \; term \; of \;} v_{SI}(m) \nonumber \\
       &:=\lim_{\epsilon \rightarrow 0} \frac{ -\Gamma}{\pi^2} \log \epsilon  \int_0^{2 \pi} \left[ \left(  t_1(0,\theta) \cdot \frac{\partial n_2(0)}{\partial s_1} \right)n_1(0) - \left(  t_1(0,\theta) \cdot \frac{\partial n_1(0)}{\partial s_1}\right)n_2(0)  \right] d \theta \label{eq:velsi}
     \end{align}
     In conclusion, the self-induced velocity field of a membrane has a logarithmic divergence, as do vortex filaments in $\mathbb{R}^3$ \cite{Sa92, ArHa65}.  The direction of this infinite velocity is determined by the integral in the above expression and this is evaluated in a later section.

\paragraph{Regularization of the infinite self-induced velocity using the LIA.}  As for filaments in $\mathbb{R}^3$, the infinite self-induced velocity can be regularized by a local induction approximation (LIA) \cite{Sa92, ArHa65}. This means that in the above evaluation one ignores the neighborhood $U_\epsilon$ of $m$ in evaluating the integral. Equivalently, fix $\epsilon$ at a finite, non-zero, small value.
The resulting leading order term in $v_{SI}(m)$ is then taken as the regularized self-induced velocity. And so
\begin{align}
       &   v_{SI,reg}(m)  \nonumber \\
       &:= \frac{ \Gamma}{\pi^2} \log \left(\frac{\delta}{\epsilon}  \right) \int_0^{2 \pi} \left[ \left(  t_1(0,\theta) \cdot \frac{\partial n_2(0)}{\partial s_1} \right)n_1(0) - \left(  t_1(0,\theta) \cdot \frac{\partial n_1(0)}{\partial s_1}\right)n_2(0)  \right] d \theta, \label{eq:vsireg} \\
        & \hspace{3in} \epsilon \; {\rm fixed,} \quad \epsilon << \delta \nonumber
     \end{align}
     Note the slightly different way in which the above is written, compared to~(\ref{eq:velsi}). The $O(1)$ term $\log \delta$ is retained to avoid attaching any spurious importance to the absolute value of $\epsilon$. Moreover, the use of the LIA and other regularizations for filaments in $\mathbb{R}^3$ have shown that, from a physical point of view, the argument for the logarithmic term is best expressed as the ratio of two length scales, $\epsilon$ and $\delta$, with $\delta$ representing a typical length scale of the vortical structure \cite{Sa92}.

\section{Hamiltonian structure for the membrane model.} In this section the Hamiltonian structure of the membrane model is presented. The Hamiltonian structure follows from the general symplectic/Poisson structure governing the dynamics of a vorticity field (in the framework of Euler's equations). Marsden and Weinstein \cite{MaWe83}, expanding on the pioneering work of Arnold, showed  formally that the appropriate phase space for the dynamics of a vorticity field specified by a vorticity two form is an infinite-dimensional manifold known as the coadjoint orbit of the dual of the Lie algebra $\mathfrak{g}^*$ (identified with the space of vorticity two-forms). The Lie algebra $\mathfrak{g}$ (identified with the space of divergence free velocity fields) is associated with the Lie group of  volume preserving diffeomorphisms of the flow domain (in general, an $n$-dimensional Riemannian manifold $M$). This manifold is equipped with a symplectic structure, known in general as the Kirillov-Kostant-Soreau symplectic structure on coadjoint orbits, and is given by
    \begin{align}
        \Omega  \left(-\mathcal{L}_u\omega, -\mathcal{L}_v \omega \right)(\omega) &= \int_{M} \omega \left(u,v \right) \mu \label{eq:mw}
    \end{align}
    In the above, the vorticity two-form $\omega$ is identified with a point on the coadjoint orbit and $(-\mathcal{L}_u \omega, -\mathcal{L}_v \omega)$---again, two-forms---are identified with tangent vectors at that point. Here $u,v \in \mathfrak{g}$ are identified with any two divergence-free vector fields on $M$ and $\mathcal{L}$ denotes the Lie derivative. More details can be found in \cite{MaWe83, AbMa78, MaRa99}.

      As applications of~(\ref{eq:mw}), Marsden and Weinstein further showed that~(\ref{eq:mw}) recovers the symplectic structure of the well-known point vortex model in $\mathbb{R}^2$ and provides a symplectic structure for the vortex filament model in $\mathbb{R}^3$.

    \paragraph{Symplectic phase space for membranes.}
Consider now the phase space $\mathcal{P}$ of membranes i.e. the images of (unparametrized) maps, modulo reparametrizations with  the same image, $\Sigma: \mathbb{R}^2 \rightarrow \mathbb{R}^4$. An element of $P$ corresponding to such a map will be denoted by $\mathfrak{p}$.

  To apply~(\ref{eq:mw}) to membranes, first evaluate the tangent vector to coadjoint orbits term for a membrane. Start with a tubular neighborhood $C$ of the geometric surface $\Sigma$ as in Appendix~\ref{strengthindep}.  It is useful to think of $C$ as a normal bundle over $\Sigma$. Choose coordinates $(l_1,l_2,s_1,s_2)$ where $(s_1,s_2)$ are coordinates for $\Sigma$ and $(l_1,l_2)$ are coordinates in the plane of normals. Choose an orthogonal frame $(n_1,n_2,t_1,t_2)$ at each point of $C$, tangent everywhere to the coordinate curves, and
   such that $(n_1,n_2)$ lie in the fiber. Without loss of generality, assume the vorticity two-form in $C$ to be $\omega=\tilde{\omega}dn_1 \wedge dn_2$, where $\tilde{\omega}$ is a constant. In other words, the vorticity two-form is the same on any plane of normals. As $C$ shrinks to $\Sigma$, $\tilde{\omega} \rightarrow \infty$  (such that the strength $\Gamma$ of $C$ remains the same)  and one obtains the membrane vorticity two-form. Write the velocity one-form on $C$  as $v^{\flat}=v_1(p)dn_1+v_2(p)dn_2+v_3(p)ds_1+v_4(p)ds_2$ where $p \equiv (l_1,l_2,s_1,s_2)$ . Use Cartan's `magic' formula \cite{MaRa99}, the closedness of $\omega$ and $v$ divergence-free, to obtain
   \begin{align*}
    \mathcal{L}_v \omega&= {\bf d}{\bf i}_v \omega= \left(v_1 \frac{\partial \tilde{\omega}}{\partial l_1} +  v_2 \frac{\partial \tilde{\omega}}{\partial l_2} \right)dn_1 \wedge dn_2 = \left(v \cdot \nabla_n \tilde{\omega} \right)dn_1 \wedge dn_2,
   \end{align*}
   where $\nabla_n$ is the gradient operator in the plane of normals. Note that (by the construction of $C$) the term in parentheses is a Dirac delta function on the boundary of $C$. In the membrane limit,
   \begin{align*}
    \mathcal{L}_v \omega&=  \left(\mid v_n \mid  \delta(m-p) \right)dn_1 \wedge dn_2, \quad m \in \mathbb{R}^4, \; p \in \Sigma
   \end{align*}
   where $v_n=\sqrt{v_1^2+v_2^2}$,  to give the identification of $\mathcal{L}_v \omega$ with a normal vector field on $\Sigma$.

   \paragraph{Remark.} The conclusion is unaltered if a more general representation of $\omega$ is chosen since in the membrane limit only the $dn_1 \wedge dn_2$ component of $\omega$, with a coefficient that is constant (though infinite!), survives. \\  \\

Substituting $\omega$ for membranes,~(\ref{eq:vortmembranes}), in the right hand side of~(\ref{eq:mw}) gives $\Gamma \int_{\Sigma}  dn_1 \wedge dn_2 \left(u,v \right) \nu$.  But since the form $dn_1 \wedge dn_2$ acts on only the normal components  at any point of $\Sigma$, the symplectic form on $\mathcal{P}$ can be finally written as
\begin{align}
    \Omega \left(X,Y \right)(\mathfrak{p})
    &= \Gamma \int_{\Sigma}  dn_1 \wedge dn_2 \left(u_n,v_n \right) \nu \label{eq:sympmemb}
\end{align}
In the above, $X,Y \in T_{\mathfrak{p}}\mathcal{P}$ and $u_n,v_n$ are the corresponding normal vector fields on $\Sigma$.

   \paragraph{Hamiltonian function.}  The Hamiltonian function on $\mathcal{P}$ is the kinetic energy of the fluid flow due to the membrane but regularized using the LIA.

            To write the fluid flow kinetic energy,
   \begin{align}
    K.E.&=\frac{1}{2} \int_{\mathbb{R}^4} V^{\flat} \wedge \ast V^{\flat}, \label{eq:ke}
   \end{align}
   as a functional of membranes it is useful to  write it first as a functional of the vorticity two-form. This requires the introduction of the equivalent of the divergence-free `vector potential' vector field $\vec{\mathcal{A}}$ in $\mathbb{R}^3$ which, it may be recalled, is related to the velocity field $\vec{v}$ in $\mathbb{R}^3$ by $\vec{v}=\nabla \times \vec{\mathcal{A}}$, and to the vorticity vector field $\vec{\omega}$ in $\mathbb{R}^3$ by Poisson's equation:
    \begin{align}
     \nabla^2 \vec{\mathcal{A}}&=-\vec{\omega} \label{eq:vecpotr3}
     \end{align}In $\mathbb{R}^4$, since vorticity  is a two-form field, the vector potential is also a two-form field satisfying similar equations as above. More precisely, the following is true in $\mathbb{R}^4$.  {\it For $V$ divergence-free, there exists a two-form $\mathcal{A}$, satisfying ${\bf d}\mathcal{A}=0$,  such that}
   \begin{align}
   - \ast {\bf d} \ast \mathcal{A}& \equiv \eth \mathcal{A}=V^{\flat} \label{eq:vecpot2form}
   \end{align}
   In the above, $\eth$ is the co-differential and acting on 2-forms in $\mathbb{R}^4$ (see \cite{AbMaRa88}), p.457). Apply the exterior derivative operator to both sides to obtain the equivalent of~(\ref{eq:vecpotr3}),
   \begin{align*}
    {\bf d} \eth \mathcal{A}&=\omega.
   \end{align*}
   The operator ${\bf d} \delta$ is the Laplace-de Rham operator without the $\eth {\bf d}$ part. The equation is then  equivalent to a Poisson equation for each component of the two-forms as shown below. Let
   \[\mathcal{A}=\sum_{i=1}^4 \sum_{j>i}^4 A_{ij} dx_i \wedge dx_j, \]
   where $A_{ij}: \mathbb{R}^4 \rightarrow \mathbb{R}$. Computing ${\bf d} \eth \mathcal{A}$
   using  ${\bf d} \mathcal{A}=0$, obtain
   \begin{align*}
    {\bf d} \eth \mathcal{A}
    &=- \sum_{i=1}^4 \sum_{j>i}^4  \nabla^2 A_{ij} dx_i \wedge dx_j,
   \end{align*}
   which is equivalent to the following set of equations in Cartesian coordinates
   \begin{align}
   \nabla^2 A_{ij}&=-\omega_{ij}, \quad j > i \label{eq:poiss4}
   \end{align}
   Existence and uniqueness of $\mathcal{A}$, for given $\omega$, therefore follows from the existence and uniqueness of solutions to Poisson's equation in $\mathbb{R}^4$. The solution to~(\ref{eq:poiss4}), using~(\ref{eq:greens}), is
   \begin{align*}
    A_{ij}(m)&= \int_{\mathbb{R}^4}  \frac{\omega_{kl}(p)}{\pi^2 (\mid r-\tilde{r} \mid^2)} \mu_p,
   \end{align*}
   where, as before, $r$ and $\tilde{r}$ are position vectors of points $m$ and $p$, respectively, and the subscript on $\mu$ denotes integration with respect to $p$.

    For a membrane, switching to the frame $(n_1,n_2,t_1,t_2)$  with origin at some $p \in \Sigma$ and using the notation
    \[\mathcal{A}(m)=A_{t_1t_2}(m)dt_1 \wedge dt_2 + A_{n_2 t_1}(m) dn_2 \wedge dt_1+ \cdots , \quad m \in \mathbb{R}^4,\]
    (with the correspondence $1 \leftrightarrow n_1,2 \leftrightarrow n_2, 3 \leftrightarrow t_1$ and $ 4 \leftrightarrow t_2$), equation~(\ref{eq:poiss4}) becomes
    \begin{align}
     \nabla^2 A_{n_1 n_2}(m)&=-\Gamma \delta(m-p), \label{eq:att}
   \end{align}
    and the right side is zero for all other components of $A$.  Note that the choice of $p$ makes no difference since the right side is always a Dirac delta function supported on $\Sigma$.  Replacing the notation $A_{n_1n_2}$ by the more meaningful notation $A_{\Sigma}$, obtain
   \begin{align}
    A_{\Sigma}(m)&=\Gamma \int_{\Sigma}  \frac{1}{\pi^2 (\mid r-\tilde{r} \mid^2)} dt_1 \wedge dt_2, \label{eq:as}
   \end{align}
   with the other components of $\mathcal{A}$ taken, without loss of generality, to be equal to zero.


   Returning to the kinetic energy integral~(\ref{eq:ke}), proceed to write it for membranes. Using the above results and integration by parts, first rewrite it as
   \begin{align*}
    K.E.&=\frac{1}{2} \int_{\mathbb{R}^4} V^{\flat} \wedge \ast \eth \mathcal{A},\\
    &= \frac{1}{2} \int_{\mathbb{R}^4} \omega \wedge \ast \mathcal{A}
   \end{align*}

    Substitute~(\ref{eq:vortmembranes}) to obtain
   \begin{align}
    K.E&=\frac{\Gamma}{2} \int_{\Sigma} A_{\Sigma}(m) \nu, \quad m \in \Sigma \nonumber \\
    &=\frac{\Gamma^2}{2 } \int_{\Sigma} \int_{\Sigma}  \frac{1}{\pi^2 (\mid r-\tilde{r} \mid^2)} \nu_p \; \nu, \quad m,p \in \Sigma \label{eq:kememb}
   \end{align}
   Of course the integral is not convergent till one regularizes the self-induced contribution. Once appropriately regularized the above integral can be viewed as a functional of $\Sigma$---since it depends only on the strength of the membrane and its geometry. And this leads to the Hamiltonian model for membranes.

   \paragraph{Regularization of the kinetic energy using the LIA.}  The regularization of the kinetic energy using the local induction approximation (LIA)  proceeds in a similar way to the regularization of  the self-induced velocity using the LIA.  In other words, one removes a neighborhood $U_\epsilon$ of $m$ in evaluating the inner integral in~(\ref{eq:kememb}). This removes the blow-up in the inner integral and the outer integral is then evaluated as a regular integral. As in the regularization of the self-induced velocity, there is no blow-up due to non-local effects.  To obtain the leading order term of the resulting kinetic energy integral for $\epsilon$ small, it suffices to evaluate the inner integral on a subdomain of $\Sigma$--in particular, the annular domain introduced previously $U:=U_\delta \backslash U_\epsilon$.
   \begin{align*}
    K.E._{reg}&:={\rm Leading \; order \; term \; in}  \frac{\Gamma^2}{2} \int_{\Sigma} \int_{U}  \frac{1}{\pi^2 (\mid r-\tilde{r} \mid^2)} \nu_p \; \nu, \quad \epsilon << \delta
\end{align*}
Using the expansions previously computed for small $s_1$, with $r=0$, obtain for the inner integral
\begin{align*}
    & \int_{U}  \frac{1}{\pi^2 (\mid \tilde{r} \mid^2)} \nu_p \\
    &= \frac{1}{\pi^2} \int_0^{2 \pi} \int_{\epsilon}^\delta   \frac{g(s_1,\theta) }{ (\mid \tilde{r}(s_1,\theta)\mid^2)} ds_1 d \theta, \\
    &= \frac{1}{\pi^2} \int_0^{2 \pi} \int_{\epsilon}^\delta   \left[\frac{1}{s_1} + O(1) \right]ds_1 d \theta, \\
    &= \frac{2}{\pi} \log c_\epsilon + O(1),
\end{align*}
 In the above,
\[ c_\epsilon:=\frac{\delta}{\epsilon} >> 1 \]
And so the regularized kinetic energy which is also the Hamiltonian function $H: \mathcal{P} \rightarrow \mathbb{R}$, is
\begin{align}
   H(\Sigma)&:=K.E._{reg}:=  \frac{\Gamma^2}{\pi} \log c_\epsilon \int_\Sigma  \; \nu, \label{eq:hamemb}
\end{align}


\paragraph{Hamiltonian vector field.} To compute the Hamiltonian vector field corresponding to~(\ref{eq:hamemb}) and~(\ref{eq:sympmemb}), the first task is to compute the variational derivative of the Hamiltonian.

Let $\delta r: \Sigma \rightarrow \mathbb{R}^4$ correspond to a variation $\delta \Sigma \in T_{\mathfrak{p}}\mathcal{P}$. The former is a field of normal  vectors on $\Sigma$.  Elements of $T_{\mathfrak{p}}^*\mathcal{P}$ are also identified with normal vectors  on $\Sigma$ and are paired with elements of $T_{\mathfrak{p}}\mathcal{P}$ using the Euclidean metric in $\mathbb{R}^4$ i.e.
\begin{align*}
    \left<u_\Sigma,\alpha_\Sigma \right>&:=\int_\Sigma <<X_u,X_\alpha >> \nu, \quad  u_\Sigma \in T_{\mathfrak{p}}\mathcal{P}, \; \alpha_\Sigma \in T_{\mathfrak{p}}^*\mathcal{P}, \\
    & =\int_{U_\Sigma \subset \mathbb{R}^2}<<X_u,X_\alpha >> ds_1 ds_2
\end{align*}
where $X_u,X_\alpha$ are the corresponding normal vector fields
(note that this pairing is degenerate if not restricted to normal fields).  The second integral implicitly assumes the existence of a partition of unity and a set of coordinate charts covering $\Sigma$. Here,  $U_\Sigma$ denotes the entire domain of integration in the coordinate charts.
The variational derivative---a normal vector field corresponding to an element of $T_{\mathfrak{p}}^*\mathcal{P}$----is then defined in the usual way:
\begin{align}
    \left<\delta\Sigma, \frac{\delta H}{\delta \Sigma} \right> & \equiv \int_{U_\Sigma \subset \mathbb{R}^2}<< \delta r,X_{\delta H / \delta \Sigma} >> ds_1 ds_2:=\lim_{\epsilon \rightarrow 0} \frac{1}{\epsilon} \left[H(\Sigma + \epsilon \delta \Sigma)-H(\Sigma) \right] \label{eq:fd}
\end{align}

Consider the local chart defined previously~(\ref{eq:chart})
\[\phi:(s_1,s_2)\mapsto (x(s_1,s_2),y(s_1,s_2),z(s_1,s_2),o(s_1,s_2))  \equiv r(s_1,s_2) \]
where, $(s_1,s_2)$ are again arc-length coordinates corresponding to an orthogonal mesh (but not necessarily the same polar mesh introduced earlier despite the use of the same symbol $s_2$).
The pullback of $\nu:=dt_{1} \wedge dt_{2} \in \Lambda^2(\mathbb{R}^4)$ under this chart is defined as
\begin{align*}
    \phi^*(dt_{1}\wedge dt_{2 })(u,v)&:=dt_{1} \wedge dt_{2} ({\bf D}\phi(u),{\bf D}\phi(v)), \quad u,v \in T_{\phi^{-1}(p)}  \mathbb{R}^2, \\
   &= (u_1 v_2 - v_1 u_2) (t_1 \cdot t_1) (t_{2} \cdot t_{2}) \\
   \Rightarrow \phi^*(dt_{1}\wedge dt_{2}) &= (t_{1} \cdot t_{1}) (t_{2} \cdot t_{2}) ds_1 \wedge ds_2,
\end{align*}
where $u_1,u_2$ are the $s_1,s_2$-components of $u$ etc. and $ds_1 \wedge ds_2 \in \Lambda^2(\mathbb{R}^2)$. The  terms $t_1 \cdot t_1$ and $t_2 \cdot t_2$ are retained in this form for now  since, as shown below, they contribute to the variations .

  Now consider the following local chart for the perturbed position of the membrane
  \begin{align*}
  \tilde{\phi}:& (s_1,s_2) \\
  & \mapsto (x+\epsilon \delta x,y+\epsilon \delta y,z+\epsilon \delta z,o+\epsilon \delta o)(s_1,s_2) \\
  & \equiv (r+ \epsilon \delta r)(s_1,s_2)
  \end{align*}
A slightly longer pullback computation then shows that
\begin{align*}
    & \tilde{\phi}^*(dt_{1}\wedge dt_{2})(u,v) \\
    &= (u_1 v_2 - u_2 v_1) \left((t_{1} \cdot t_1) (t_2 \cdot t_2) + \epsilon((\delta t_1 \cdot t_1)(t_2 \cdot t_2) +(\delta t_2 \cdot t_2)(t_1 \cdot t_1)  )+ O(\epsilon^2) \right), \\
    &=\left((t_{1} \cdot t_1) (t_2 \cdot t_2) + \epsilon((\delta t_1 \cdot t_1)(t_2 \cdot t_2) +(\delta t_2 \cdot t_2)(t_1 \cdot t_1)  )+ O(\epsilon^2) \right) ds_1 \wedge ds_2,
\end{align*}
The functional derivative $\delta H/ \delta \Sigma$, is identified with a normal vector field on  $\Sigma$. From~(\ref{eq:fd}), obtain
\begin{align}
    \left<\delta \Sigma, \frac{\delta H}{\delta \Sigma} \right>&:= \int_{U_\Sigma \subset \mathbb{R}^2} \left< \left<\delta r(\phi(s_1,s_2)), X_{\delta H / \delta \Sigma}(\phi(s_1,s_2)) \right> \right>\; ds_1 ds_2  \nonumber \\
    &:= \frac{\Gamma^2}{\pi} \log c_\epsilon \frac{1}{\epsilon} \lim_{\epsilon \rightarrow 0} \left[ \int_{\Sigma + \epsilon \delta \Sigma} dt_{1}  \wedge dt_{2}- \int_{\Sigma} dt_{1} \wedge dt_{2}  \right], \nonumber \\
    &=\int_{U_\Sigma \subset \mathbb{R}^2} ((\delta t_1 \cdot t_1)(t_2 \cdot t_2) +(\delta t_2 \cdot t_2)(t_1 \cdot t_1)  ) \; ds_1 ds_2, \nonumber \\
    &=\int_{U_\Sigma \subset \mathbb{R}^2} (\delta t_1 \cdot t_1 +\delta t_2 \cdot t_2 ) \; ds_1 ds_2, \label{eq:tdelt}
\end{align}
To manipulate the above integral, consider the following 1-form in on $\Sigma$, \[(\delta r \cdot t_1) dt_2 \]
Since $\Sigma$ is compact and boundaryless, from Stokes theorem
\begin{align}
    \int_\Sigma {\bf d} \left( (\delta r  \cdot t_1) dt_2 \right)&=0 \label{eq:stokes}
\end{align}
Now, in a local chart,
\begin{align}
    {\bf d} \left( (\delta r  \cdot t_1)  dt_2 \right)&= \frac{\partial ( (\delta r  \cdot t_1)(\phi (s_1,s_2)))}{\partial s_1} dt_1  \wedge dt_2, \nonumber \\
    \Rightarrow \int_{\Sigma } {\bf d} \left( (\delta r  \cdot t_1) dt_2 \right) &= \int_{U_\Sigma \subset \mathbb{R}^2} \left( \frac{\partial \delta r }{\partial s_1} \cdot t_1 + \frac{\partial t_1}{\partial s_1} \cdot \delta r  \right) (\phi (s_1,s_2)) ds_1  ds_2, \label{eq:comm}
\end{align}
Using ${t}_{1}=\partial{r} / \partial s_1,t_2 =\partial{r}/ \partial s_2$
and making the usual assumption of variations commuting with differentiation, i.e.
\begin{align*}
\delta t_1 &=\frac{\partial{\delta r}}{\partial s_1}, \quad \delta t_2 =\frac{\partial{\delta r}}{\partial s_2}
\end{align*} it follows  from~(\ref{eq:stokes}) and~(\ref{eq:comm}) that
\begin{align*}
    \int_{U_\Sigma \subset \mathbb{R}^2} (\delta t_1 \cdot t_1  ) \; ds_1 ds_2&=  - \int_{U_\Sigma \subset \mathbb{R}^2} (\frac{\partial t_1}{\partial s_1} \cdot \delta r  ) \; ds_1 ds_2, \\
    \int_{U_\Sigma \subset \mathbb{R}^2} (\delta t_2 \cdot t_2 ) \; ds_1 ds_2&=  - \int_{U_\Sigma \subset \mathbb{R}^2} (\frac{\partial t_2}{\partial s_2} \cdot \delta r  ) \; ds_1 ds_2
\end{align*}
Substituting back in~(\ref{eq:tdelt}), finally obtain
\begin{align*}
    X_{\delta H / \delta \Sigma}(p)&=-\frac{\Gamma^2}{\pi} \log c_\epsilon \left(\frac{\partial t_1}{\partial s_1} + \frac{\partial t_2}{\partial s_2} \right)(p)
\end{align*}
Since, $X_{\delta H / \delta \Sigma}$ has non-trivial components only in the normal directions, we can write
\[\frac{\partial t_1}{\partial s_1} + \frac{\partial t_2}{\partial s_2}(p)=\left(\left(\frac{\partial t_1}{\partial s_1} + \frac{\partial t_2}{\partial s_2} \right) \cdot n_1, \left(\frac{\partial t_1}{\partial s_1} + \frac{\partial t_2}{\partial s_2} \right) \cdot n_2 \right) (p), \]
where $n_1(p),n_2(p)$ denote the basis vectors in the plane of normals at $p$.
Using $t_i(p) \cdot n_j(p)=0$ $\forall p$, the components on the right (in the $n_1(p)-n_2(p)$ plane) can also be written as \[- \left( \frac{\partial n_1  }{\partial s_1} \cdot t_1 + \frac{\partial n_1  }{\partial s_2} \cdot t_2\right) (p)  \]
etc. Letting,
\begin{align}
\mathfrak{K}(p)&:=-\frac{1}{2} \left(\frac{\partial n_1 }{\partial s_1} \cdot t_1 + \frac{\partial n_1 }{\partial s_2} \cdot t_2, \frac{\partial n_2  }{\partial s_1} \cdot t_1 + \frac{\partial n_2 }{\partial s_2} \cdot t_2 \right)(p) \label{eq:mcv}
\end{align}
 it follows that
\begin{align*}
    X_{\delta H / \delta \Sigma}(p)&=-\frac{2\Gamma^2}{\pi} \log c_\epsilon \mathfrak{K}(p)
\end{align*}

The Hamiltonian vector field $X_H$ is obtained from the general form of Hamilton's equation on symplectic manifolds \cite{MaRa99}:
\begin{align*}
    \Omega(X_H,\delta \Sigma)&= \left<\delta \Sigma, \frac{\delta H}{\delta \Sigma} \right>
\end{align*}
Using~(\ref{eq:sympmemb}) and the above derived result for the variational derivative, this equation becomes
\begin{align*}
   & \Gamma \int_{\Sigma}  dn_1 \wedge dn_2 \left(X_H,  \delta r \right) \nu=\int_{U_\Sigma \subset \mathbb{R}^2} \left< \left<\delta r, -\frac{2\Gamma^2}{\pi} \log c_\epsilon \mathfrak{K} \right> \right>\; ds_1 ds_2, \\
    & \Rightarrow   \Gamma \int_{U_\Sigma \subset \mathbb{R}^2}  \left< \left<\delta r, \left(X_H \cdot n_1 \right)n_2-    \left(X_H \cdot n_2 \right)n_1\right> \right>\; ds_1 ds_2  \\
    & \hspace{2.5in}=\int_{U_\Sigma \subset \mathbb{R}^2} \left< \left<\delta r, -\frac{2\Gamma^2}{\pi} \log c_\epsilon \mathfrak{K} \right> \right>\; ds_1 ds_2,
\end{align*}
Since $\delta r$ has non-trivial components only in normal directions, this gives
\begin{align*}
    X_H&= R_{-\pi/2} \cdot \frac{2\Gamma}{\pi} \log c_\epsilon \mathfrak{K},
\end{align*}
where $R_{-\pi/2}=\left(\begin{array}{cc} 0 & -1 \\ 1 & 0 \end{array} \right)$, and Hamilton's equations for the membrane evolution is
\begin{align}
    \frac{\partial \Sigma}{\partial t}&=X_H=R_{-\pi/2} \cdot \frac{2\Gamma}{\pi} \log c_\epsilon \mathfrak{K} \label{eq:membevo}
\end{align}

\section{The mean curvature vector field.}
In this short section, it will be shown that $\mathfrak{K}$ in~(\ref{eq:mcv}) is the mean curvature vector field of $\Sigma$. Moreover, it will be shown that the right hand side of~(\ref{eq:membevo}) is the same as the regularized self-induced velocity field~(\ref{eq:vsireg}) obtained from a direct computation.

  Recall that for a 2D surface $\Sigma$ in $\mathbb{R}^3$, the second fundamental form at $p \in \Sigma$ is the quadratic form
  \begin{align*}
    \mathfrak{S}(V)(p)&:=-\left< \left< {\bf D}n(p) \cdot V), V \right> \right>, \quad V \in T_p \Sigma,
\end{align*}
where $(n,t_1,t_2)$ are the unit vectors of a moving orthogonal frame and the unit normal vector is viewed as a map $n:\Sigma \rightarrow S^2$ so that ${\bf D}n(p):T_p \Sigma (\simeq \mathbb{R}^2) \rightarrow T_{n(p)}S^2 (\simeq \mathbb{R}^2)$ is the derivative map \cite{Ca94, Th79}. The mean curvature $\kappa$ is then defined as
\[\kappa(p)=\frac{1}{2} \left(\mathfrak{S}(t_1)+ \mathfrak{S}(t_2) \right)(p) \]
For a 2D surface $\Sigma$ in $\mathbb{R}^4$, there is a second fundamental form associated with each of the normal directions, $n_i$, of the moving frame $(n_1,n_2,t_1,t_2)$.
 \begin{align*}
    \mathfrak{S}_1(V)(p)&:=-\left< \left< {\bf D}n_1(p) \cdot V), V \right> \right>, \quad \mathfrak{S}_2(V)(p):=-\left< \left< {\bf D}n_2(p) \cdot V), V \right> \right>, \\
    & \hspace{3.5in} \quad V \in T_p \Sigma,
\end{align*}
 The {\it mean curvature vector} $\mathfrak{K}$ is then defined as (see, for example, \cite{Ch71})
\begin{align}
\mathfrak{K}(p)&=\sum_{i=1}^2\frac{1}{2} \left(\mathfrak{S}_i(t_1)+ \mathfrak{S}_i(t_2)\right)(p)n_i(p)  \label{eq:mcvgen}
 \end{align}
and one checks that~(\ref{eq:mcv}) is the same as the above.

  To show that the right hand side of~(\ref{eq:membevo}) is the same as the regularized self-induced velocity field~(\ref{eq:vsireg}) is equivalent to showing that
\begin{align}
   \frac{1}{2 \pi} \int_0^{2 \pi} t_1(0,\theta) \cdot  \frac{\partial n_1(0)}{\partial s_1} d \theta&= -\mathfrak{K}(m) \cdot n_1(0)  \label{eq:equiv1}\\
   \frac{1}{2 \pi} \int_0^{2 \pi} t_1(0,\theta) \cdot  \frac{\partial n_2(0)}{\partial s_1} d \theta&= -\mathfrak{K}(m) \cdot n_2(0) \label{eq:equiv2}
\end{align}
(reverting to the notation used in the self-induced velocity field section).  However, this is a straightforward exercise and for the case of $\mathbb{R}^3$ can actually be found as an exercise problem in, for example, Thorpe's textbook \cite{Th79}(Exercises 12.16 and 12.17). Begin by noting that the integrands on the left are nothing but $-\mathfrak{S}_i(t_1(0,\theta))(m)$, $i=1,2$. Since $t_1=t_1(0,0)$ and $t_2=t_1(0,\pi/2)$,  write $t_1(0,\theta)=\cos \theta t_1+ \sin \theta t_2$. Substituting for $t_1(0,\theta)$ and performing the integrations, one obtains that the integrals are equal to $(-1/2) (\mathfrak{S}_i(t_1)+\mathfrak{S}_i(t_2))(m)$.  Using~(\ref{eq:mcvgen}), proves~(\ref{eq:equiv1}) and~(\ref{eq:equiv2}).

 It is interesting to contrast the dynamics of membranes, as given by~(\ref{eq:membevo}), with the dynamics of the vortex surfaces considered in \cite{Li98,JiLi06} where the surfaces move {\it parallel} to the mean curvature vector.

\section{The dynamics of $\omega \wedge \omega$ and an application to Ertel's vorticity theorem in $\mathbb{R}^3$.}
As an application of vortex dynamics in $\mathbb{R}^4$, it will be shown in this section that Ertel's vorticity theorem in $\mathbb{R}^3$, for the constant density case, is simply a special case of the dynamics of the four-form $\omega \wedge \omega$. A few general ideas related to this four-form are discussed first.

   \paragraph{Basics of $\omega \wedge \omega$.} All four-forms are identically zero in $\mathbb{R}^3$ and lower dimensions.
The four-form $\omega \wedge \omega$ in $\mathbb{R}^4$ is given by
 \begin{align}
    \omega^2&= 2\left( \omega_{12} \omega_{34} - \omega_{13} \omega_{24} + \omega_{14} \omega_{23} \right) dx_1 \wedge dx_2 \wedge dx_3 \wedge dx_4  \label{eq:alpha}
 \end{align}
 It is an exact form by definition
\begin{align}
	\omega \wedge \omega &= {\bf d}v^{\flat} \wedge \omega= {\bf d} \left(v^{\flat} \wedge \omega \right) \label{eq:helform}
\end{align}
and by Stokes' theorem it follows that
\begin{align}
	\int_{D \subset \mathbb{R}^4} \omega \wedge \omega&= \int_{\partial D} i^*\left(v^{\flat} \wedge \omega \right)
\end{align}
implying $\int_{\mathbb{R}^4} \omega \wedge \omega=0$ if $\omega$ has compact support or decays sufficiently rapidly at infinity.

    It is also straightforward to see from~(\ref{eq:lpv}), using derivation properties of the operators involved, that
\begin{align}
	\frac{\partial (\omega \wedge \omega)}{\partial t}+ \mathcal{L}_v (\omega \wedge \omega)&=0 \label{eq:om2}
\end{align}
Further, writing
\begin{align}
	\omega \wedge \omega&= f_{2} \mu \equiv f_{2} \wedge \mu,
\end{align}
where $\mu$ is the canonical volume form on $\mathbb{R}^4$, it follows using derivation properties and the fact that $v$ is divergence-free, i.e $\mathcal{L}_v \mu=0$, that
\begin{align}
	\frac{\partial f_{2}}{\partial t}+ \mathcal{L}_v  f_{2} &=0, \label{eq:omf2}
\end{align}
A consequence of the general results of Khesin and Chekanov \cite{KhCh89} for an $n$-dimensional Riemannian manifold is that the Euler's equations in $\mathbb{R}^4$ admit an infinite number of integrals of the form $\int_{\mathbb{R}^4} f(f_2) \mu$. Arnold and Khesin  \cite{ArKh98} call $f_2$ the `vorticity function' and discuss several of its properties, see also the discussion in \cite{TuYa93}. The dynamics of $\omega \wedge \omega$ is also studied in the numerical and theoretical work of \cite{GoWaShNaSu07}. Interestingly, the earliest work known to the author that studies equation~(\ref{eq:omf2}), and generalizations, is a relatively lesser-known work of Hill \cite{Hi1885}, also cited by Truesdell \cite{Tr54} (p.173).  Indeed, Hill examined Euler's equations on an $n$-dimensional Euclidean space.

  \paragraph{Application to Ertel's vorticity theorem.} First, consider the case when the flow is with respect to a non-rotating, inertial frame. Consider $f(x,y,z,t): \mathbb{R}^3 \rightarrow \mathbb{R}$. Let $w$ be a divergence-free velocity field in $\mathbb{R}^3$ satisfying Euler's equation for a constant-density fluid flow
  \begin{align*}
    \frac{\partial w}{\partial t}+ w \cdot \nabla w&=- \nabla\frac{ p}{\rho},
 \end{align*}
 where the $\nabla$ operators are in $\mathbb{R}^3$.
 The following vector field in $\mathbb{R}^4$
  \[v:=(w,f), \]
 is therefore also divergence-free. For $v$ to satisfy Euler's equation it follows that $f$ must satisfy
  \begin{align*}
    \frac{\partial f}{\partial t} + w \cdot \nabla f&= 0
  \end{align*}
  Euler's equation, written for the velocity 1-form $v^{\flat}$ \cite{ArKh98},
  \begin{align}
    \frac{\partial v^{\flat}}{\partial t}+ \mathcal{L}_v v^{\flat}&={\bf d}g, \quad g: \mathbb{R}^4 \rightarrow \mathbb{R} \label{eq:1formeq}
  \end{align}
  where the ${\bf d}$ operator is in $\mathbb{R}^4$, is then equivalent to the equations
  \begin{align*}
    \frac{\partial w}{\partial t}+ w \cdot \nabla w&= -\nabla \left( \frac{w^2 + f^2}{2}\right) + \nabla g = \nabla  \left( -\frac{w^2 + f^2}{2} + g\right), \\
    \frac{\partial f}{\partial t} + w \cdot \nabla f&= \frac{\partial g}{\partial o}
  \end{align*}
  Euler's equation for $w$ and the conservation law for $f$ are both recovered by choosing $g(x,y,z,o)= (w^2 + f^2)/2 - p/\rho$.

  The six components of the vorticity two-form are then
  \[\omega=(\omega_i,{\bf d} f), \]
  where $\omega_i$ is the vorticity two-form ${\bf d}w^{\flat}$. Denoting by $X_{\omega_i}$ the vector field in $\mathbb{R}^3$ corresponding to $\omega_i$, i.e. $X_{\omega_i}=(\ast_{\mathbb{R}^3} \omega_i)^{\sharp} \equiv (\omega_{23},-\omega_{13},\omega_{12})$, obtain
  \[\omega \wedge \omega= 2(X_{\omega_i}\cdot \nabla f) dx \wedge dy \wedge dz \wedge do  \]
  And so from~(\ref{eq:om2}) and~(\ref{eq:omf2}), one obtains Ertel's Theorem for the potential vorticity $X_{\omega_i}\cdot \nabla f$ in a non-rotating frame \cite{Pe87},
  \begin{align*}
    \frac{D}{Dt} \left(X_{\omega_i}\cdot \nabla f \right)&=0,
  \end{align*}
  where $D/Dt \equiv \partial / \partial t + w \cdot \nabla$.

    To demonstrate the result when the flow is with respect to a constantly rotating frame---the usual setting of Ertel's theorem---the Lie-Poisson formulation of Euler's equation in a rotating frame in $\mathbb{R}^n$ needs to be briefly discussed first. Recall \cite{ArKh98, MaWe83} that in non-rotating, inertial frames in $\mathbb{R}^n$, Euler's equations can be written as Hamilton's equations associated with a Lie-Poisson bracket on $\mathfrak{g}^*$---the dual of the Lie algebra of the Lie group of volume preserving diffeomorphisms of $\mathbb{R}^n$----as follows
    \begin{align}
        \left<\frac{\partial [v^{\flat}]}{\partial t}, \frac{\delta F}{\delta [v^{\flat}]} \right>&=-\left<v^{\flat}, \left[ \frac{\delta F}{\delta [v^{\flat}]}, \frac{\delta H}{\delta [v^{\flat}]}\right] \right> \label{eq:lpnr}
    \end{align}
    Here, $v \in \mathfrak{g}$ is a divergence-free velocity field in the inertial frame and $[v^{\flat}] \equiv {\bf d}v^{\flat} \in \mathfrak{g}^*$ is the equivalence class of 1-forms generated by $v^{\flat} \sim v^{\flat} +{\bf d}h$, $h: \mathbb{R}^n \rightarrow \mathbb{R}$, and identified with the vorticity two-form of $v$. The right hand side of the above equation is the Lie-Poisson bracket on $\mathfrak{g}^*$. The pairing $\left<\;,\; \right>$ between elements of $\mathfrak{g}$ and $\mathfrak{g}^*$ is given by $\left<[v^{\flat}],w \right>:=\int_{\mathbb{R}^n} v^{\flat}(w) \; \mu$ and is non-degenerate due to the $L^2$-orthogonality of gradient fields and divergence-free fields. The Hamiltonian function $H: \mathfrak{g}^* \rightarrow \mathbb{R}$ is the flow kinetic energy $H([v^{\flat}])=(1/2) \int_{\mathbb{R}^n} v^{\flat} \wedge \ast v^{\flat}$ and is a well-defined function, i.e. independent of the choice of the representative element of $[v^{\flat}]$ used in the integral, because of the $L^2$-orthogonality. Computing the functional derivative gives $\delta H /\delta [v^{\flat}]=v$ and using Lie bracket $[ \; ]$ properties one obtains Euler's equations as
    \[ \frac{\partial [v^{\flat}]}{\partial t}=- \mathcal{L}_v [v^{\flat}],\]
    from which one obtains both~(\ref{eq:1formeq}) and~(\ref{eq:lpv}).

         To apply this formalism to a constantly rotating frame in $\mathbb{R}^n$, let $v_r \in \mathfrak{g}$ be a divergence-free velocity field of the flow observed in the rotating frame. Formally, $\mathfrak{g}$ can be viewed as the Lie algebra of the Lie group of volume preserving diffeomorphisms of $\mathbb{R}^n$ that go to elements of $\operatorname{SO}(n)$ at infinity. More specifically, each volume preserving diffeomorphism in this group has associated with it a unique element of $\operatorname{SO}(n)$. Equivalently, each $v_r$ has associated with it a unique element of $\mathfrak{so}(n)$---the Lie algebra of $\operatorname{SO}(n)$. The uniqueness follows simply by implementing the infinity condition in the inertial frame which is that the volume preserving diffeomorphism goes to the identity map or, equivalently, the velocity field goes to zero. Note that the uniqueness of course does not extend the other way.

         The dual of the Lie algebra, $\mathfrak{g}^*$, is identified with elements $[v_r^{\flat}] \equiv {\bf d}v_r^{\flat} \equiv \omega_r$, where $\omega_r$ is the vorticity two-form of the flow observed in the rotating frame. The equivalence class is again generated by gradient fields in the rotating frame that go to zero at infinity. The pairing is then defined as \[\left<[v_r^{\flat}],u_r \right>:=\int_{\mathbb{R}^n} (v_r^{\flat} + \ast (\ast \Omega_v \wedge l^{\flat}))(u_r + (\ast (\ast \Omega_u \wedge l^{\flat}))^{\sharp})\; \mu,\] where $\Omega_v,\Omega_u \in \mathfrak{so}(n)$ are constant two-forms in $\mathbb{R}^n$. The velocity fields $ (\ast (\ast \Omega_u \wedge l^{\flat}))^{\sharp})$ etc are rigid body angular velocity fields in $\mathbb{R}^n$, where $l$ is position vector from the center of the frame, and note that these are also divergence-free. One checks that this pairing is also non-degenerate due to the $L^2$-orthogonality of gradient fields and divergence-free fields. The Hamiltonian function is again the fluid flow kinetic energy written as $H([v_r^{\flat}]):=(1/2) \int_{\mathbb{R}^n} (v_r^{\flat} + \ast (\ast \Omega_v \wedge l^{\flat}))\wedge \ast (v_r^{\flat} + (\ast (\ast \Omega_v \wedge l^{\flat})))\; \mu$, invariant under the change $v_r^{\flat} \rightarrow v_f^{\flat} + {\bf d}h$. The functional derivative is computed as $\delta H /\delta [v_r^{\flat}]=v_r +  (\ast (\ast \Omega_v \wedge l^{\flat}))^{\sharp}$.

             The Lie-Poisson form of Euler's equation in the rotating frame is then given by~(\ref{eq:lpnr}) with $v^{\flat}$ replaced by $v_r^{\flat}$. One obtains
 \begin{align}
  \frac{\partial [v_r^{\flat}]}{\partial t}&=- \mathcal{L}_{v_r} [v_r^{\flat}+(\ast (\ast \Omega_v \wedge l^{\flat})) ] \nonumber
  \end{align}
  The equations for the velocity 1-form $v_r^{\flat}$ and the vorticity two-form $\omega_r$ are then given, respectively, by
  \begin{align}
    \frac{\partial v_r^{\flat}}{\partial t}+ \mathcal{L}_{v_r} (v_r^{\flat}+(\ast (\ast \Omega_v \wedge l^{\flat})))&={\bf d}g, \label{eq:1formeqrot} \\
    \frac{\partial \omega_r}{\partial t}+ \mathcal{L}_{v_r} (\omega_r+{\bf d}(\ast (\ast \Omega_v \wedge l^{\flat})))&=0, \nonumber
  \end{align}
  In $\mathbb{R}^3$, the above vorticity equation reduces to the familiar equation for the vorticity vector field of the flow observed in a rotating frame \cite{Pe87}. Since $\Omega_v$ is invariant in time, for the vorticity two-form
  \[\omega:=\omega_r+{\bf d}(\ast (\ast \Omega_v \wedge l^{\flat})), \]
  one readily obtains the evolution equation
  \begin{align}
     \frac{\partial (\omega \wedge \omega)}{\partial t}+ \mathcal{L}_{v_r} (\omega \wedge \omega)&=0, \quad  \frac{\partial f_2}{\partial t}+ \mathcal{L}_{v_r} f_2=0,\label{eq:lpoworot}
  \end{align}
  where, as before, $\omega \wedge \omega=f_2 \mu$.

    To obtain Ertel's theorem via the above equation in $\mathbb{R}^4$, consider a constantly rotating frame $K_3$ in $\mathbb{R}^3$. This can be viewed as a constantly rotating frame $K_4$ in $\mathbb{R}^4$ with the $o$-axis fixed. The constant angular velocity two-form of $K_4$ is $\Omega=\Omega_z dx \wedge dy - \Omega_y dx \wedge dz + \Omega_z dy \wedge dz$. Consider the vector field $v_r$ in $\mathbb{R}^4$ given by
    \[ v_r:=(w_r,f),\]
    where $w_r$ is a divergence-free velocity field in $K_3$ and, as before, $f(x,y,z,t): \mathbb{R}^3 \rightarrow \mathbb{R}$, so that $v_r$ is also divergence-free. One checks that
    \begin{align}
     v_r+(\ast (\ast \Omega \wedge l^{\flat}))^{\sharp}&:=(w_r+(\ast (\ast \Omega \wedge l_3^{\flat}))_3^{\sharp},f), \nonumber \\
     &=(w_r+X_{\Omega} \times l_3,f), \nonumber \\
     \Rightarrow \omega&= (\omega_r+2 \Omega,{\bf d}f), \nonumber \\
     \Rightarrow \omega \wedge \omega&= \left((X_{\omega_r}+2 X_\Omega) \cdot \nabla f \right)\mu, \label{eq:oworot}
     \end{align}
    where $l_3$ is the projection of the position vector $l$ onto $\mathbb{R}^3$, $X_\Omega=(\Omega_x,\Omega_y,\Omega_z)$ and the subscript for the parentheses on the right in the first equation denotes that all exterior algebraic operations are performed in $\mathbb{R}^3$.

      The system of equations obtained from~(\ref{eq:1formeqrot}) is then
      \begin{align*}
        \frac{\partial w_r}{\partial t}+ w_r \cdot \nabla w_r+2X_\Omega \times w_r&= \nabla \left( -\frac{w_r^2 + f^2}{2}-(X_\Omega \times l_3)\cdot w_r  +  g \right), \\
    \frac{\partial f}{\partial t} + w_r \cdot \nabla f&= \frac{\partial g}{\partial o}
      \end{align*}
       Euler's equation for $w_r$ in $K_3$ and the conservation law for $f$ are both recovered by choosing $g(x,y,z,o)= (w_r^2 + f^2)/2 +(X_\Omega \times l_3)\cdot w_r + \Phi - p/\rho$, where the gradient of $\Phi$ is the centripetal acceleration term in $K_3$. Finally, from~(\ref{eq:lpoworot}) and~(\ref{eq:oworot}), one obtains Ertel's theorem for the potential vorticity $(X_{\omega_r}+2 X_\Omega)\cdot \nabla f$ in a rotating frame \cite{Pe87}
       \begin{align*}
        \frac{D}{Dt} \left((X_{\omega_r}+2 X_\Omega) \cdot \nabla f \right)&=0,
       \end{align*}
       where $D/DT \equiv \partial / \partial t + w_r \cdot \nabla$.

  A derivation similar in spirit to this is carried out in Subin and Vedan \cite{SuVe04} (section 6) but, seemingly, not with the objective of obtaining Ertel's theorem. The  authors choose the $o$-coordinate as time and $f=w^2/2$ and obtain a  final equation which is not quite Ertel's theorem. The relation of Ertel's theorem to symmetries in Lagrangian and Hamiltonian systems and to symplecticity has been discussed previously in the literature---for example, \cite{Ri81, Sa82, BrHyRe05}. See also the recent work of Gibbon and Holm \cite{GiHo11}.

\section{Summary and future directions.} This paper examines the vortex  dynamics of Euler's equations for a constant density fluid flow in $\mathbb{R}^4$. The focus is on singular Dirac delta distributions of vorticity supported on two-dimensional surfaces termed membranes. The self-induced velocity field of a membrane has a logartihmic divergence. A regularization done via the local induction approximation (LIA) then shows that the  regularized velocity field is proportional to the mean curvature vector field of the membrane rotated by 90$^{\circ}$ in the plane of normals. A Hamiltonian structure for the regularized self-induced motion of the membrane is also presented. Finally, the dynamics of the four-form $\omega \wedge \omega$ is examined and it is shown that Ertel's vorticity theorem in $\mathbb{R}^3$, for the constant density case, is a special case of this dynamics.

        An interesting future direction to pursue is how Hasimoto's transformation \cite{Ha72} can be generalized and applied to equation~(\ref{eq:membevo}), and to investigate the relation of the transformed nonlinear PDE to the nonlinear Schr\"{o}dinger equation. Constructing membrane models for the dynamics of $\omega \wedge \omega$ and potential vorticity is another direction. As one possibility, models where the potential vorticity is a Dirac delta function supported on points could be constructed. In other words, a point potential vortex model. However, in general, such models will not provide a closed system of evolution equations for the points---unlike for point vortices in  $\mathbb{R}^2$---since potential vorticity is governed by the evolution of both the vorticity field and the passive scalar. Hamiltonian formalisms also exist for incompressible, non-constant density and for compressible fluid flows as also for ideal magnetohydrodyanmic flows \cite{MoGr80, HoMaRa98}. Investigating the role of $\omega \wedge \omega$, or an appropriately generalized quantity, in such flows in higher-dimensional spaces is another interesting direction.

\paragraph{Acknowledgements.} The author is grateful to Darryl D. Holm for conversations and to Ravi N. Banavar for pointing him to the textbook by Thorpe.

\newpage
\appendix

\section{Proof that the strength of a membrane is a well-defined constant.} \label{strengthindep}

\paragraph{Proof that Definition~(\ref{strength}) defines the strength of the membrane.}
  Recall that for a filament $\mathcal{C}$ in $\mathbb{R}^3$, its strength is defined by first considering a de-singularized vortex tube. This is a geometric construct and is a {\it tubular neighborhood} $C \subset \mathbb{R}^3$ of (the curve) $\mathcal{C}$ (see, for example, Ana De Silva \cite{Si01} for a discussion of tubular neighborhoods) such that at every point in $\partial C$, $i^*_{\partial C} \omega=0$ where $i_{\partial C}: \partial C \rightarrow C$ is the inclusion map. In other words, vortex lines passing through points in $\partial C$ are everywhere tangent to $\partial C$. Moreover, it is assumed that there exists another tubular neighborhood $\tilde{C} \supset C$ such that there is no vorticity in $\tilde{C} \backslash C$. The strength of the vortex tube is defined as $\Gamma:=\int_A i_A^* \omega$, where $A \subset C$ is again a two-dimensional disc-like surface whose boundary $\partial A$ is a non-contractible loop  on $\partial C$ and $i_A:A \rightarrow \mathbb{R}^3$ is the inclusion map. Now for any two curves $\partial A_1$ and $\partial A_2$  apply Stokes theorem to the part of the tube bounded by $A_1$, $A_2$ and the part of $\partial C$ that is bounded by $\partial A_i$. Using the fact that  $\omega$  is a closed form in $C$ and $i^*_{\partial C} \omega=0$  delivers the result $\int_{A_1} i_{A_1}^* \omega=\int_{A_2} i_{A_2}^* \omega$.  In other words,
  the strength of the tube is independent of the choice of  $\partial A$. This invariance persists
  as the tube approaches the singular limit of the filament.

    For membranes, the procedure is similar. However, some additional steps are needed due to the extra spatial dimension involved. By the classical tubular neighborhood theorem  (see again \cite{Si01}), there exists a tubular neighborhood $C \subset \mathbb{R}^4$ of (the surface) $\Sigma$. Analogous to vortex lines being tangent to the boundary of a  vortex tube in $\mathbb{R}^3$, assume that $C$ is such that at each point of $\partial C$---a three-dimensional `manifold'--- $i_{\partial C}^* \omega$=0, where $i_{\partial C}:\partial C \rightarrow C$ is the inclusion. This is equivalent to saying that each point of $\partial C$ there passes a (unique) vortex surface $\mathcal{S}$ such that $i_{\mathcal{S}}^* \omega=0$ and $\mathcal{S} \subset \partial C$. Again, assume the existence of a neighborhood $\tilde{C} \supset C$ such that the vorticity two-form is zero in $\tilde{C} \backslash C$.  Now since $C$ is four-dimensional and ${\bf d}\omega$ is three-dimensional, Stokes theorem cannot be directly applied to $C$ or to any four-dimensional submanifold.

    Consider therefore a tubular neighborhood $B_l$ of a  smooth closed curve $l  \in \Sigma$. $B_l \subset C $ is three-dimensional and $\partial B_l \subset \partial C$. Let $i_B: B_l \rightarrow C$, then ${\bf d}i_B^* \omega=0$ since ${\bf d}\omega=0$ everywhere in $C$ and the exterior derivative operation on forms commutes with pullbacks by mappings i.e  ${\bf d}i_B^* \omega=i_B^* {\bf d} \omega$. $B_l$ is  like a vortex tube in $\mathbb{R}^3$ but the ambient space is $\mathbb{R}^4$. Define the strength of $B_l$ accordingly:
    \begin{align}
    \Gamma_{l}:&= \int_A i_A^* \omega,\label{eq:gammal}
     \end{align}with $A$ as before ($\partial A$ a non-contractible loop on $\partial B_l$) and $i_A:A \rightarrow \mathbb{R}^4$ is the inclusion. As for a vortex tube in $\mathbb{R}^3$, the next step is to show that $\Gamma_l$ is a well-defined constant invariant with the choice of $\partial A$. Denote by $B_{1,2}$ the part of  $B_l$  with piecewise smooth boundary   $\partial B_{1,2}=A_1 \cup A_2 \cup \partial S$, where $\partial S$ is the part of  $ \partial B_l$ bounded by $\partial A_i$. The next step is to apply Stokes theorem for $i_B^* \omega$ in $B_{1,2}$.
    \begin{align*}
        \int_{B_{1,2}} {\bf d} i_B^* \omega&=\int_{\partial S} \tilde{i}^*_{\partial S} i^*_B \omega + \int_{A_1} \tilde{i}^*_{A_1} i^*_B \omega + \int_{A_2} \tilde{i}^*_{A_2} i^*_B \omega,
 \end{align*}
 where $\tilde{i}_{\partial S}: \partial S \rightarrow B_l$, $\tilde{i}_{A_1}: A_1 \rightarrow B_l$ and $\tilde{i}_{A_2}: A_2 \rightarrow B_l$. As shown previously, the left hand side is zero. Now, the pullbacks in the integrands on the right are the pullbacks by the composition maps $i_B \circ \tilde{i}_{\partial S}$ etc. which are the same as the inclusions $i_{\partial S}: \partial S \rightarrow C$ etc. From the description of $\partial S$ above, it is obvious that $i_{\partial S}$ is the same as $i_{\partial B_l}: \partial B_l \rightarrow C$.  And so the three boundary integrals above can be written as
 \[\int_{\partial S} i^*_{\partial B_l}  \omega + \int_{A_1} i^*_{A_1}  \omega + \int_{A_2} i^*_{A_2}  \omega \]
 But, since $\partial B_l \subset \partial C$,  $i_{\partial B_l}$ is nothing but a restriction of the (previously defined) inclusion $i_{\partial C}: \partial C \rightarrow C$. It follows that $i^*_{\partial B_l} \omega= \hat{i}^*_{\partial B_l} i^*_{\partial C} \omega=0$, since $i^*_{\partial C} \omega=0$, where $\hat{i}_{\partial B_l}: \partial B_l \rightarrow \partial C$ is the inclusion. And so the first boundary integral vanishes, the second and third represent $\Gamma_l$, as defined by~(\ref{eq:gammal}), on $A_1$ and $A_2$, respectively (keep in mind that the interior of $C$ is everywhere like $\mathbb{R}^4$ so that $i_{A_1}$ and $i_{A_2}$ are really inclusions in $\mathbb{R}^4$) and, after taking into account the orientations of the $A_i$s it follows that $\Gamma_l$ is the same for $\partial A_1$ and $\partial A_2$.

     The above shows that the strength of $B_l$ is a constant for a given smooth closed  curve $l \in \Sigma$. The final step is to show that this strength is the same for the $B_l$ of {\it any} smooth closed curve in  $\Sigma$.  For this,  consider two smooth closed curves $l_1,l_2 \in \Sigma$ which {\it coincide on a finite segment of the curves}, denoted by $l_{ab}$, and their tubular neighborhoods $B_{l_1}$ and $B_{l_2}$, respectively. It follows that $B_{l_1} \cap B_{l_2} \neq \phi$. More importantly, $B_{l_{ab}}:=C_{\mid_{l_{ab}}} \subset B_{l_1} \cap B_{l_2}$ is non-empty. Considering a non-contractible $\partial A \in \partial B_{l_{ab}}$, it is then straightforward to show, using the constancy of $\Gamma_{l_1}$ and $\Gamma_{l_2}$,  that $\Gamma_{l_1}=\Gamma_{l_2}$. This geometric construction of two smooth closed curves that coincide on a finite segment can obviously also be used to connect any two {\it disconnected} curves in $\Sigma$---since $\Sigma$ is assumed to be arcwise-connected---by a third smooth closed curve which shares finite identical segments with each of the disconnected curves. A transitivity argument then shows that the $B_l$s of the two disconnected curves must have the same strength $\Gamma_l$. And this suffices to show that the $B_l$ of {\it any} smooth closed curve in $\Sigma$ has the same strength $\Gamma_l$. Passing to the singular limit then proves the result $\blacksquare$

\newpage

\bibliographystyle{new}    

\end{document}